\documentclass{PoS}

\def\mathswitchr#1{\relax\ifmmode{\mathrm{#1}}\else$\mathrm{#1}$\fi}
\def\mathswitch#1{\relax\ifmmode#1\else$#1$\fi}
\newcommand{\PW}{\mathswitchr W}
\newcommand{\PZ}{\mathswitchr Z}
\newcommand{\Pl}{\mathswitch l}
\newcommand{\Pp}{\mathswitchr p}

\newcommand{\X}{X}
\newcommand{\rT}{{\mathrm{T}}}
\newcommand{\MW}{\mathswitch {M_\PW}}
\newcommand{\ri}{{\mathrm{i}}}
\newcommand{\GW}{\mathswitch {\Gamma_\PW}}
\newcommand{\EW}{{\mathrm{EW}}}

\def\citere#1{\mbox{Ref.~\cite{#1}}}

\def\de{\delta}

\title{Electroweak precision for W+jet production}

\ShortTitle{Electroweak precision for W+jet production}

\author{Ansgar Denner\\
        Paul Scherrer Institut, W\"urenlingen und Villigen,\\ 
        Ch-5232 Villigen PSI, Switzerland\\
        E-mail: \email{ansgar.denner@psi.ch}}

\author{Stefan Dittmaier\\
        Albert-Ludwigs-Universit\"at Freiburg, Physikalisches Institut, \\
        D-79104 Freiburg, Germany\\
        E-mail: \email{stefan.dittmaier@physik.uni-freiburg.de}}

\author{Tobias Kasprzik\\
        Karlsruhe Institute of Technology (KIT), 
        Institut f\"ur Theoretische Teilchenphysik, \\        
        D-76128 Karlsruhe, Germany\\
        E-mail: \email{kasprzik@particle.uni-karlsruhe.de}}

\author{\speaker{Alexander M\"uck}\\
        RWTH Aachen, Institut f\"ur Theoretische Physik E,\\
        D-52056 Aachen, Germany\\
        E-mail: \email{mueck@physik.rwth-aachen.de}}

\abstract{
In this talk we discuss the next-to-leading-order electroweak (EW) corrections 
to \mbox{W-boson + jet} hadroproduction~\cite{Denner:2009gj} and compare the
full result to a simple approximation assuming factorization
of EW and QCD corrections for the charged-current Drell--Yan process. 
The \PW-boson resonance is treated
consistently using the complex-mass scheme, and all off-shell effects
are taken into account. The corresponding next-to-leading-order QCD
corrections have also been recalculated. All the results are
implemented in a flexible Monte Carlo code. Selected numerical results 
for this Standard Model benchmark process are presented for the LHC.
The comparison of our result to an approximation based on the EW corrections
to \PW-boson production without additional jets is a step towards a
better understanding of the interplay between QCD and EW effects
for \PW-boson production in general.
}

\FullConference{RADCOR 2009 - 9th International Symposium on Radiative Corrections (Applications of Quantum Field Theory to Phenomenology) ,\\
                 October 25 - 30 2009\\
                 Ascona, Switzerland}

\begin{document}

\section{Introduction}

The production of electroweak (EW) \PW\ and \PZ\ bosons with subsequent
leptonic decays is one of the cleanest and most frequent Standard Model (SM)
processes at the Tevatron and
the LHC. The charged-current Drell--Yan process allows for a precision
measurement of the \PW-boson mass and width, can deliver important 
constraints in the fit of the parton distribution functions, may 
serve as a luminosity monitor at the LHC, and offers the possibility
to search for new charged $\PW{}^\prime$ gauge bosons. For more details
we refer the reader for example to \citere{Gerber:2007xk} and references 
therein.

At hadron colliders, the EW gauge bosons are (almost) always
produced together with additional QCD radiation. The production
cross section of \PW\ bosons in association with a hard, visible jet,
\begin{equation}
\Pp\Pp/\Pp\bar\Pp \to \PW  + \mathrm{jet} \to \Pl\nu_\Pl + \mathrm{jet} +\X,
\end{equation}
is still large. The jet recoil can lead to strongly boosted
\PW\ bosons, i.e.\ to events with high-$p_{\rT}$ charged leptons and/or 
neutrinos. Hence, $\PW+\mathrm{jet(s)}$ production is not only a 
SM candle process, it is also an important background for a large class 
of new physics searches based on missing transverse momentum. Moreover, 
the process offers the possibility for precision tests concerning jet 
dynamics in QCD.

To match the prospects and importance of this process class, an
excellent theoretical accuracy has already been achieved for the 
prediction of inclusive $\PW$-boson production including NNLO calculations,
resummation, parton-shower matching, NLO EW corrections, and leading 
higher-order corrections. The production of $\PW$ bosons in association 
with jets is now known in NLO QCD up to 3 jets~\cite{Berger:2009ep}. 
An extensive list of references can be found in \citere{Denner:2009gj}.

So far, the EW corrections in the SM have been assessed for 
$\PW+1\,\mathrm{jet}$ production in an on-shell approximation where 
the \PW\ boson is treated as a stable external 
particle~\cite{Kuhn:2007qc}. For \PW\ bosons
at large transverse momentum, i.e.\ at large centre-of-mass energy,
this is a good approximation since the EW corrections are dominated by
large universal Sudakov logarithms.

In this work, we summarize a calculation of the NLO EW
corrections for the physical final state in \PW-boson hadroproduction,
i.e.\ $\Pp\Pp/\Pp\bar\Pp \to \Pl\nu_\Pl + \mathrm{jet} +\X$,
described in full detail in \citere{Denner:2009gj}. 
In contrast to the on-shell approximation, all off-shell
effects due to the finite width of the \PW\ boson are included. Moreover,
we can incorporate the experimental event selection based on the charged-lepton 
momentum and the missing transverse momentum of the neutrino in 
our fully flexible Monte Carlo code
which is able to calculate binned distributions for all physically
relevant $\PW+1\,\mathrm{jet}$ observables. 
Our calculation, introduced in Section~\ref{se:calculation}, is completely 
generic in the sense that it can
predict observables which are dominated by \PW\ bosons close to their
mass shell as well as observables for which the exchanged \PW\ boson
is far off-shell. Moreover, we have recalculated the NLO QCD
corrections at $\mathcal{O}(\alpha^2_\mathrm{s} \alpha^2)$, 
supporting a phase-space dependent choice for the
factorization and renormalization scales. Selected results are discussed
in Section~\ref{se:results}.

The calculation of the EW corrections to \PW\ 
production in association with a hard jet is also a step towards a
better understanding of the interplay between QCD and EW corrections
for \PW\ production in general. More specifically, our calculation 
allows to test the approximation which assumes factorization for EW and QCD 
corrections in \PW\ production in a simple but well controlled setup: 
Calculating the
EW corrections to single-\PW\ production and taking into the account the 
emission of the additional jet in a subsequent step is compared to 
our calculation, which constitutes a part of the full NNLO mixed EW/QCD 
corrections for single-\PW\ production, in Section~\ref{se:single_W}.
The understanding of the interplay between QCD and EW effects---including 
a full treatment of off-shell W~bosons---is mandatory to
match the envisaged experimental accuracy for the \PW-mass measurement
at the Tevatron and the LHC.

\section{The Calculation}
\label{se:calculation}

In this section we highlight specific aspects of the calculation 
which are particularly important for the presented corrections and
which are not part of the standard framework for NLO calculations. For 
an extensive discussion of the calculational setup we refer the reader 
to \citere{Denner:2009gj}.

The potentially resonant \PW\ bosons require a proper inclusion of the
finite gauge-boson width in the propagators. We use the complex-mass
scheme~\cite{Denner:2005fg}. In this approach the W-boson mass (as well as
the Z-boson mass) is consistently considered as a complex quantity, 
\begin{equation}
\mu_{\PW}^2 = \MW^2 -\ri \MW \GW \, ,
\end{equation}
defined as the location of the propagator pole in the complex plane, where
\MW\ is the conventional real mass and $\GW$ denotes the \PW-boson width.
This leads to complex couplings and, in particular, a complex weak mixing 
angle. The underlying (real) Lagrangian does not change since the 
introduced width is compensated  by adding a corresponding complex 
counterterm. The scheme fully respects all relations that follow from 
gauge invariance. 

The experimental event definition for final-state muons usually selects
so-called ``bare'' muons which are measured without any special treatment of
collinear bremsstrahlung photons. Technically, the two collinear particles are
not recombined into a single pseudo-particle and the observable is 
not collinear safe. Therefore, the KLN theorem does not apply and the 
corresponding EW corrections include terms which are enhanced by
logarithms of the (small) muon mass. The enhanced corrections are
phenomenologically relevant and cannot be calculated by the standard subtraction
methods which assume collinear safety. Accordingly, we use an extended
dipole subtraction method~\cite{Dittmaier:2008md} which has been specifically 
designed to deal with non-collinear-safe observables. The logarithms are
extracted analytically and we can still work with matrix elements in the
massless muon approximation.

To form collinear-safe quantities, QCD partons and also photons
have to be recombined into a single jet
if they are sufficiently collinear. However, the recombination induces a 
problem if the bremsstrahlung photon and a gluon are accidentally
collinear. In this case, soft gluons can still pass
the jet selection due to the recombination procedure. Hence, a soft-gluon
divergence is induced that would be canceled by the virtual QCD 
corrections to $\PW+\mathrm{photon}$ production. To avoid the singularity,
one has to distinguish $\PW+\mathrm{photon}$ and $\PW+\mathrm{jet}$ 
production by means of a
more precise event definition employing a cut on the maximal energy or
transverse-momentum fraction of a photon inside a given jet. However,
this procedure spoils the collinear safety of the event definition in
partonic processes with final-state quarks. Using
again the subtraction formalism~\cite{Dittmaier:2008md} to extract the
problematic collinear terms, the appearance of an unphysical quark-mass 
logarithm in the final result signals the necessity to include
non-perturbative physics to properly describe the emission of a photon
by a quark. The relevant collinear physics can be factorized from the 
underlying hard process and can be cast into a process-independent 
quark-to-photon fragmentation function~\cite{Glover:1993xc}, which has 
been measured at LEP in photon+jet events~\cite{Buskulic:1995au}. We
employ this fragmentation function to achieve both, a realistic event 
selection and a theoretically consistent result.

To reach the accuracy of $\mathcal{O}(\alpha_\mathrm{s} \alpha^3)$
throughout the calculation we have also included the photon-induced
partonic processes and the respective NLO QCD corrections. Also
non-trivial interference terms between EW and QCD diagrams
within the real corrections have been included at this order. However, these
contributions are phenomenologically irrelevant and will not be
discussed in this talk.

\section{Results}
\label{se:results}

We define $\PW+1\,\mathrm{jet}$ events by requiring a jet and a charged lepton 
with transverse momentum $p_{\rT} > 25\,$GeV as well as missing transverse
momentum larger than $25\,$GeV. The jet and the lepton have to be central with a 
rapidity smaller then 2.5 in absolute value. The details of the event selection
as well as the numerical input values for the calculation can be found in 
\citere{Denner:2009gj}. All results are presented for the LHC running at 
$14\,$TeV.

\begin{figure}
\includegraphics[width=7.25cm]{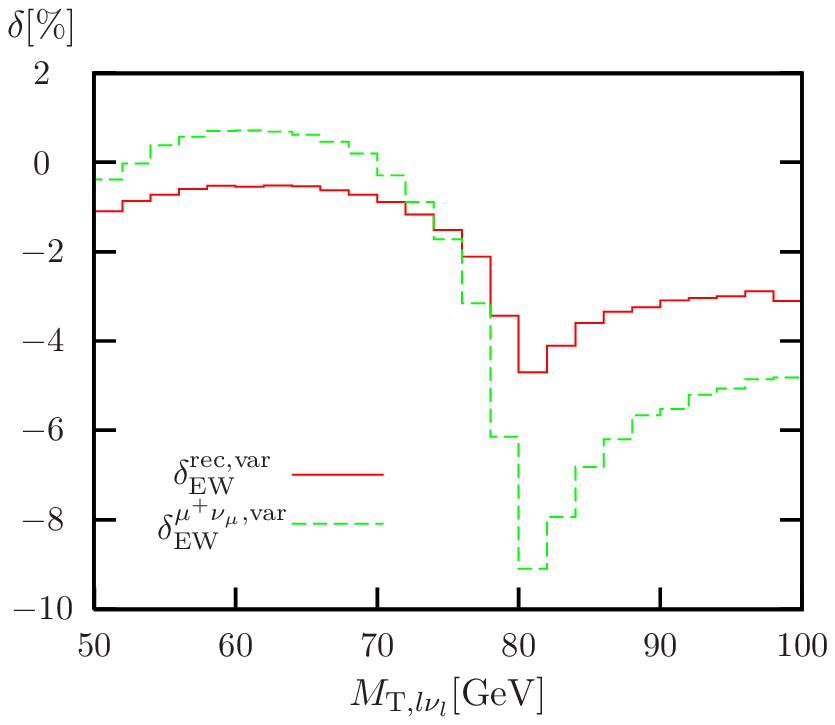}
\hfill
\includegraphics[width=7.25cm]{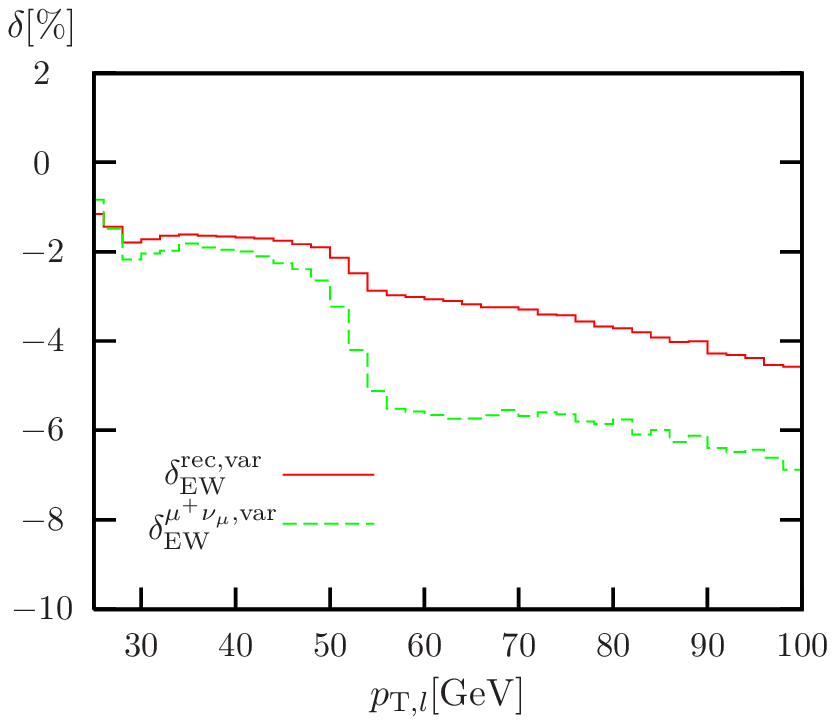}
\caption{\label{fi:results}
EW corrections to the transverse-mass distribution of the leptons (left) and to the 
transverse-momentum distribution of the charged lepton (right) at the LHC. See text 
for details.}
\end{figure}

For the inclusive cross section, we find negative percent-level EW 
corrections. When we focus on events in the tails of the transverse-momentum 
distributions of the charged-lepton $p_{\rT,\Pl}$ or the jet 
$p_{\rT,\mathrm{jet}}$ (or the transverse-mass distribution of the 
final-state leptons $M_{\rT,\Pl\nu_\Pl}$) we observe the well-known 
universal Sudakov enhancement of EW corrections in the high-energy regime.
For example, at $p_{\rT} = 1\,$TeV for the leading jet, the EW corrections 
rise to $-25\%$. In the Sudakov regime, where the on-shell result is 
a good approximation, the transverse-momentum distribution for the 
leading jet agrees at the percent level with the previous on-shell 
results~\cite{Kuhn:2007qc}. 

For all results 
in this talk we employ a variable scale choice (var) which reflects the 
kinematics of the process and has been chosen to stabilize the QCD 
corrections (see \citere{Denner:2009gj}). Concerning the QCD corrections, 
we only briefly note that a veto against a second hard QCD jet has to be 
used to carefully define the $\PW+1\,\mathrm{jet}$ observable, in particular 
for the $p_{\rT,\mathrm{jet}}$ distribution. Otherwise, the differential 
cross section is completely dominated by QCD dijet production, where a 
quark jet radiates a \PW\ boson, i.e.\ by a completely different process which 
is not related to a generic NLO contribution. 

In contrast to the integrated cross sections, the transverse-mass 
distribution is quite sensitive to the specific
treatment of final-state photons, in particular close to the Jacobian peak of
the distribution at $M_{\rT,\Pl\nu_\Pl}\sim \MW$, where the correction for
bare muons, $\de_{\EW}^{\mu^+\nu_{\mu}\,,\mathrm{var}}$, reaches almost 
$-10\%$  (see left panel of Figure~\ref{fi:results}). As expected, the 
corrections for bare muons are larger than the corrections with lepton--photon recombination, 
$\de_{\EW}^{\mathrm{rec}\,,\mathrm{var}}$, since photons, being radiated 
collinearly to the charged lepton, carry away transverse momentum.
The region around the Jacobian peak, $M_{\rT,\Pl \nu_\Pl}\sim\MW$, is of 
particular interest for the precise determination of the \PW-boson mass. 

The EW corrections for $M_{\rT,\Pl \nu_\Pl}\sim\MW$ near the Jacobian peak
resemble the corrections 
for the inclusive \PW-boson sample for which no additional jet is required
(see, e.g., Figure~2 in \citere{Brensing:2007qm}). The fact that an additional 
jet due to QCD initial-state radiation does not have a large effect on 
the EW corrections indicates that EW and QCD effects approximately factorize 
for the transverse-mass distribution close to \MW. For the transverse momentum
of the charged lepton, $p_{\rT,\Pl}$ (see right panel of Figure~\ref{fi:results}), 
the EW corrections are quite different from the single-\PW\ results 
(Figure~1 in \citere{Brensing:2007qm})
and
we discuss the question of factorization for this observable in detail in 
the next section.

\section{Testing Factorization of QCD and EW Corrections in W Production}
\label{se:single_W}

In this section, we compare the EW corrections to $\PW+1\,\mathrm{jet}$ with a 
simple approximation based on the EW corrections for \PW\ production without any
additional jet activity. 
This comparison can shed some light on the  important question how the available
EW and QCD corrections can be combined to obtain the most accurate predictions for
the charged-current Drell-Yan process while a full calculation for the
mixed $\mathcal{O}(\alpha \alpha_s)$ corrections is missing. We test the assumption
that the EW and QCD corrections factorize, motivated by
the fact that QCD does not couple to the leptonic final state and that the EW
corrections are dominated by collinear final-state radiation from the charged lepton. In
general, in this approximation a given observable can be first calculated
including the EW corrections for \PW\ production but ignoring all QCD effects. Then all the 
relevant known QCD corrections can be applied to this result, e.g.\ fixed-order and/or 
resummed corrections and/or parton-shower evolution of the final state. 
For a recent discussion combining several tools and estimating
the theoretical error of different approximations see~\citere{Balossini:2009sa}.

\begin{figure}
\includegraphics[width=7.25cm]{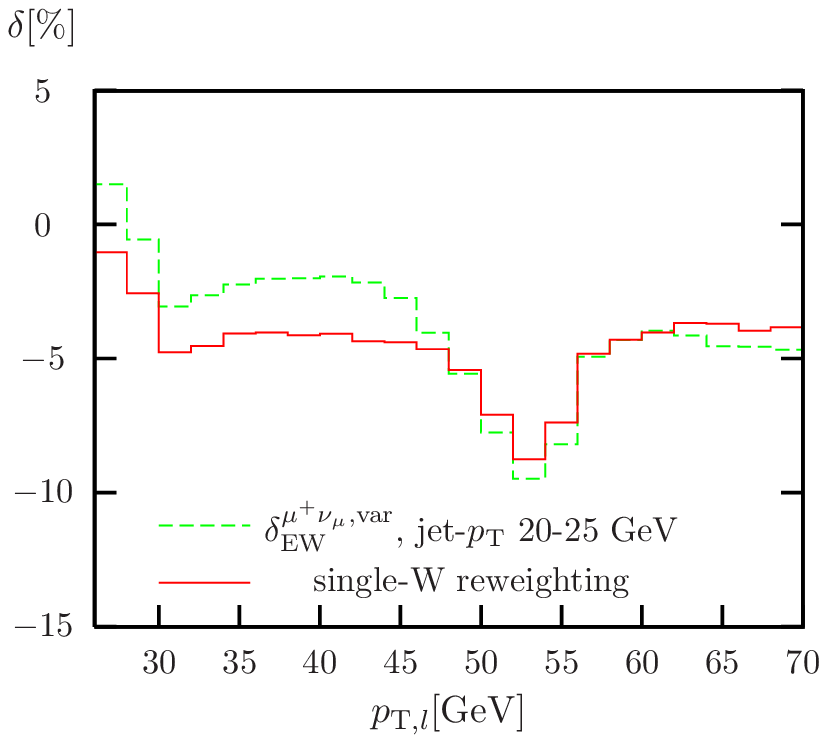}
\hfill
\includegraphics[width=7.25cm]{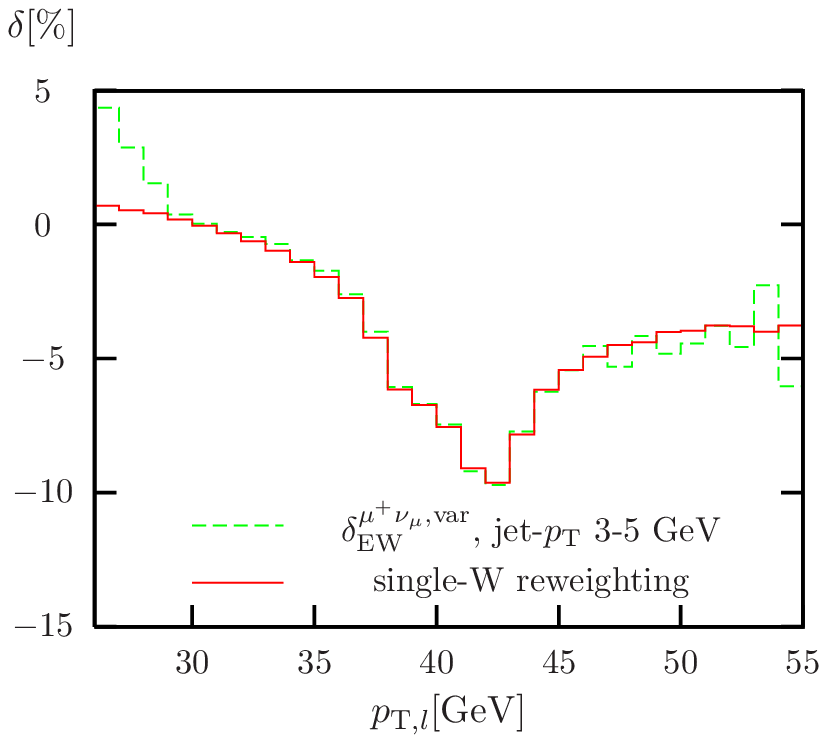}
\caption{\label{fi:single_W_comparison}
EW corrections for the $p_{\rT,\Pl}$ distribution obtained from the full calculation 
and the factorization approximation by reweighting the single-\PW\ result, as explained 
in the text. The left plot shows the corrections for events
with $20\,$GeV $<p_{\rT,\mathrm{jet}}<25\,$GeV, the right plot for events
with $3\,$GeV $<p_{\rT,\mathrm{jet}}<5\,$GeV.}
\end{figure}

Here we follow this prescription for leptonic observables 
in events where QCD radiation produces an additional jet. For the EW 
corrections to the underlying single-\PW\ production, 
we use the results from \citere{Brensing:2007qm} tuned to our 
$\PW+1\,\mathrm{jet}$ setup employing the
complex-mass scheme. To describe the QCD radiation resulting in a jet our approach is very
modest: we simply use the tree-level $\PW+1\,\mathrm{jet}$ matrix elements to
describe the first QCD emission. Hence, we do not seek for most accurate 
predictions. But on the other hand, we can test the assumed 
factorization because we can compare to the complete EW corrections to 
$\PW+1\,\mathrm{jet}$ production which include all possible 
cross-talk between QCD emission and EW effects at the level
of $\mathcal{O}(\alpha \alpha_s)$ corrections to the Drell--Yan
process for this specific contribution.

Technically, the comparison is realized as follows: We first calculate a 
tree-level $\PW+1\,\mathrm{jet}$ event. Then we reweight this event according to
the corresponding EW corrected prediction for the underlying \PW\ production.
The reweighting factor is obtained by boosting the event into the
\PW-boson rest frame and looking up the EW correction for single-\PW\ production 
in the histogram for the leptonic observable under consideration, e.g.\
$p_{\rT,\Pl}$ for the results discussed in the following.

Here, we focus on the transverse momentum $p_{\rT,\Pl}$ of the 
charged lepton where the direct sensitivity of the observable to the jet 
recoil clearly obscures or may even spoil the factorization approximation. 
Indeed, the approximation fails for events including hard jets which are 
present in our default setup. The more 
complicated kinematical situation cannot be captured by the simple reweighting
procedure advertised above. However, this is not the kinematical region
where the combination of EW and QCD effects is most needed for the \PW-mass
measurement, for which events with small QCD recoil are selected. 

In Figure~\ref{fi:single_W_comparison} (left), we show the full EW corrections and the
result from the reweighting approximation for a restricted class of events
with $20\,$GeV $<p_{\rT,\mathrm{jet}}<25\,$GeV for the transverse momentum of the jet.
Around $p_{\rT,\Pl}\sim 55\,$GeV, where the EW corrections show a dip due to the 
remnant of the Jacobian peak of the cross section in this region,
the factorization approximation works quite well. However, for smaller $p_{\rT,\Pl}$
the approximation underestimates the full result by an amount which is as big as the
correction itself. In this region, final-state configurations of decaying 
on-shell \PW\ bosons often fail to pass the missing $p_{\rT}$ cut for the given
$p_{\rT,\mathrm{jet}}$ and $p_{\rT,\Pl}$. In the full calculation, events with real photon emission populate
the region suppressed at tree level and reduce the negative EW corrections. This, 
of course, is an effect the reweighting procedure cannot account for. The harder the jets in the
events the more such kinematical effects related to cuts are relevant for the total EW 
corrections, and it is not surprising that the factorization approximation fails for 
the inclusive $p_{\rT,\Pl}$ distribution, where 
different regions of the distribution are dominated by events with different 
$p_{\rT,\mathrm{jet}}$. 

On the other hand, for events with little QCD activity, corresponding to low 
$p_{\rT,\mathrm{jet}}$ in our simple approach, the factorization approximation can be expected to 
work. The tree-level approximation in QCD for the $\PW+1\,\mathrm{jet}$ cross 
section, of course, breaks down at low $p_{\rT,\mathrm{jet}}$. However, the test of 
factorization may still be performed since only the EW corrections are relevant, not the
cross section itself. As expected, for events with $3\,$GeV $<p_{\rT,\mathrm{jet}}<5\,$GeV, the
approximation is almost exact, as shown in Figure~\ref{fi:single_W_comparison} 
(right). The region subject to kinematical complications at the edge of the 
distribution is very small.

Similar considerations apply for the $M_{\rT,\Pl\nu_\Pl}$ distribution.
However, since $M_{\rT,\Pl\nu_\Pl}$ is not sensitive to initial-state
radiation, the region $M_{\rT,\Pl\nu_\Pl}\sim \MW$ is not strongly affected by the 
discussed kinematical effects. Therefore, the factorization close 
to the Jacobian peak is visible already by directly comparing the single-\PW\ and the
$\PW+1\,\mathrm{jet}$ results for the EW corrections. Using the proposed 
approximation allows to reproduce the full $\PW+1\,\mathrm{jet}$ result even more closely.

\section{Conclusion}
\label{se:conclusion}

We have extended the theoretical effort for 
the precise prediction for \PW-boson production at the Tevatron and the
LHC by an important step: We have presented the first calculation of the full 
electroweak NLO corrections for \PW-boson hadroproduction in association 
with a hard jet where all off-shell effects are taken into account in the 
leptonic \PW-boson decay, i.e.\ we have studied final states with a jet, a
charged lepton, and missing transverse momentum at NLO in the EW coupling
constant within the SM. All results are implemented in a flexible 
Monte Carlo code which can model the experimental event definition at the 
NLO parton level. Comparing our calculation with a simple approximation 
indicates that EW corrections to \PW\ production approximately 
factorize from the underlying QCD dynamics for certain observables in limited 
kinematical regions.

\end{document}